# Comment on "Frequency response and origin of the spin-1/2 photoluminescence-detected magnetic resonance in a π-conjugated polymer"

C. G. Yang[1], E. Ehrenfreund[2], and Z. V. Vardeny[1]

[1] Department of Physics, University of Utah, Salt Lake City, Utah 84112, USA
[2] Department of Physics and Solid State Institute, Technion-Israel Institute of Technology, Haifa 32000, Israel

*In a recent paper Segal et al. [1] attempted to explain the dynamics of spin-½ photoluminescence detected magnetic resonance (PLDMR) in films of a π-conjugated polymer, namely a soluble derivative of poly(phenylene-vinylene) [MEH-PPV] using a model (dubbed TPQ), in which the PLDMR is due to spin-dependent triplet-polaron interactions that reduce the polarons density and consequent quenching of singlet excitons. We studied the full PLDMR and photoinduced absorption (PA) dynamics of MEH-PPV films as a function of microwave power at various temperatures. We show, firstly, that the TPQ model is incompatible with the full frequency dependent spin-½ PLDMR response; secondly, it is not in agreement with the spin-1 PLDMR temperature dependence; thirdly, it predicts a much shorter triplet exciton lifetime than that obtained experimentally; and fourthly, that is in contradiction with the temperature dependencies of spin-½ PLDMR and triplet exciton PA. In contrast, an alternative model, namely the spin dependent recombination of polarons, is capable of explaining the whole body of experimental results, and in particular the PLDMR dynamics.*

Segal et al. [1] reported the dynamics of spin-½ photoluminescence detected magnetic resonance (PLDMR) and photoinduced absorption (PA) in films of the archetypal π-conjugated polymer, namely poly[2-methoxy-5-(2-ethylhexyloxy)-1,4-phnylenevinylene] (MEH-PPV) at 20 K. They claimed that the frequency response of the resonance is consistent with a triplet-polaron quenching (TPQ) model, in which the resonance is mediated by spin-dependent interactions between photogenerated triplet excitons (TE) and polaron pairs in the film [1]. We measured the *full* dynamics of both spin-½ and spin-1 PLDMR resonances in MEH-PPV films as a function of microwave power at various temperatures. In this Comment we show that the TPQ model fails to explain the data. We checked that an alternative model, namely the spin dependent recombination (SDR) of polarons [2], readily accounts for the results.

In the TPQ model [1] the photogenerated TE interact with spin-paired polarons by collisions that enhance the polaron recombination. The spin dependence of the TE-polaron annihilation process in this model is a result of *spin conservation*. The spin subsystem of a colliding spin-1 TE and spin-½ polaron comprises of six spin-states of equal probability, in which only two have spin ½. After the TE is annihilated, the excited polaron with spin-½ remains; therefore spin conservation disallows 2/3 of TE-polaron collisions. Resonant spin-½ conditions induce rapid transitions between the spin-½ sublevels so that all TE-polaron collisions become allowed [1]. The TPQ model for explaining the spin-½ PLDMR resonance is therefore viable under two important conditions, *which may be readily checked by the experiment*. Firstly, in addition to photogenerated polaron density, a substantial density of long-lived photogenerated TE should also exist in the film; and secondly, the TE spin-lattice relaxation time should be longer than the TE-polaron collision time, so that their spin-state is not randomized before colliding with the paired polarons; in other words, the TE spin-lattice relaxation rate should be relatively small. Measuring spin-1 and spin-½ PLDMR resonance dynamics, and PA of polarons and TE at various temperatures scrutinize these two conditions as reported here.

In addition, Segal et al. [1] also calculated the microwave frequency ($f_M$) response dynamics of the spin-½ PLDMR resonance based on the TPQ model (Eq. (26) in ref. [1]); and used it to fit the experimental PLDMR dynamics. Alas, only the magnitude ($|\Delta PL|$) of the spin-½ PLDMR frequency response was measured in [1], where $|\Delta PL| = [((\Delta PL_I)^2 + (\Delta PL_Q)^2]^{1/2}$, and $\Delta PL_I$ and $\Delta PL_Q$ are the in-phase and quadrature components of the change $\Delta PL$ in the photoluminescence (*PL*) at resonance. Thus unfortunately, important information on the PLDMR dynamics was missed; with rather severe consequences. In our PLDMR experiments we measured *both* $\Delta PL$ components vs. $f_M$ at various microwave powers to ensure that we register the *full* dynamics of the spin-½ PLDMR resonance. When the *full* PLDMR dynamics is unraveled, then it becomes obvious that the TPQ model fails to reproduce the data. This is important since the PLDMR dynamics can disclose the underlying mechanism for the resonance, and thus the failure to reproduce the data shows that the TPQ model is irrelevant for explaining the PLDMR in π-conjugated polymers.

The PA and PLDMR measurements were conducted at various temperatures on a MEH-PPV film drop-casted from a toluene solution that was mounted in a high Q microwave cavity. The polymer film was excited with an Ar$^+$ laser at 488 nm with an intensity of ~ 500 mW/cm$^2$ subjected to spin-½ (*H* = 1070 Gauss) and spin-1 (at 'half field', *H* = 370 Gauss) resonance conditions at ~3 GHz (*S*-band) microwave frequency [2]. For PA, an incandescent light source was used, and the changes $\Delta T$ in the transmission *T* caused by the laser illumination at various modulation frequency, $f_L$ were measured using phase-sensitive technique. Both the in-phase and quadrature components of $\Delta T$ were routinely recorded. For PLDMR, we measured the changes $\Delta PL$ in *PL* caused by the magnetic resonance, where the microwave intensity was modulated at various frequencies, $f_M$; again both the in-phase and quadrature $\Delta PL$ components were measured, where the phase, ϕ was set with respect to the microwave modulation. In addition, the PLDMR was studied under variable microwave power conditions, *P* in the range 2.5 to 100 mW.

Fig. 1 shows the spin-½ PLDMR response vs. $f_M$ at 20 K and *P* = 80 mW for the two $\Delta PL$ components; $|\Delta PL|$ and the phase ϕ vs. $f_M$ were also calculated and shown for completeness. The measured $|\Delta PL(f_M)|$ response is quite similar to the response obtained in [1] indicating that the polymer sample and resonance conditions in the two laboratories are very similar. However, by measuring the microwave modulation frequency response of both $\Delta PL$ components, an unexpected surprise is unraveled; this was completely over-

looked in [1]. As seen in Fig. 1, the in-phase component $\Delta PL_I(f_M)$ *changes sign* at a frequency $f_0$ of about 30 kHz before decaying away at higher frequencies. Importantly, this response is unique for the in-phase component; the quadrature component retains its sign within the same experimental frequency range. The phase response $\phi(f_M)$ shows the sign change in $\Delta PL_I(f_M)$ more clearly; it crosses the value $\phi = \pi/2$ at ~ 30 kHz and continues to decrease thereof as $f_M$ increases. We checked that this curious PLDMR dynamic behavior is not an artifact of the measuring set-up by changing the microwave power, $P$. Fig. 1 (inset) shows the dependence of $f_0$ with $P$. We found that $f_0$ increases with $P$, and thus cannot be an artifact. Moreover, $f_0$ changes when varying the laser excitation intensity, or when films of different polymers and semiconductors were measured. This bizarre PLDMR $f_M$-response cannot be detected when measuring only $|\Delta PL(f_M)|$; thus the true PLDMR dynamics was completely missed in [1]. Moreover, it cannot be explained by a simple one- or two-oscillators response as introduced in [1] for $|\Delta PL|$ dynamics. A much more profound understanding of PLDMR dynamic response must be involved for explaining the astonishing $\Delta PL_I(f_M)$ dynamics and its dependence on $P$ [3].

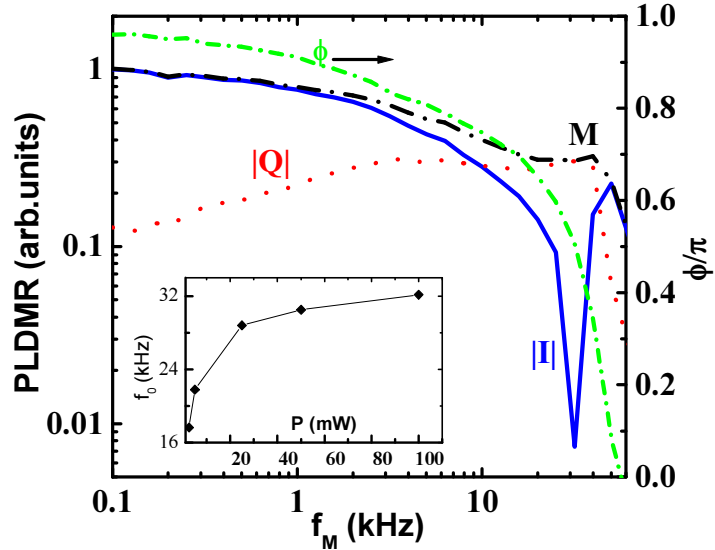

**Fig. 1:** The spin-½ PLDMR, $\Delta P$ vs. the microwave modulation frequency, $f_M$ for a MEH-PPV film at 20 K. The in-phase (I, blue; solid line) and quadrature (Q, red; dotted line) PLDMR components are shown separately, as well as the magnitude $|\Delta PL|$ (M, black; dash-dotted line) and the phase (Φ, green; short dash-dotted line; right scale). Note the zero crossing of the in-phase $\Delta PL$ component at $f_0 \approx$ 30 kHz, for microwave power $P$=50 mW. The inset shows the dependence of $f_0$ on $P$.

We first attempt to explain the surprising PLDMR dynamics using the TPQ model introduced in [1]. The PLDMR vs. $f_M$ response was fitted in [1] using the following two-oscillators equation for the complex $\Delta PL(f_M)$ response {Eq. (26) in ref. [1]};

$$\Delta PL(f_M)/PL = c_M(i\omega + z_{M1})/[(i\omega + p_{M1})(i\omega + p_{M2})], \quad (1)$$

where $c_M$ is a scaling factor, $\omega = 2\pi f_M$, and $z_M$, $p_{M1}$ and $p_{M2}$ are some effective decay rates, which were determined by the TPQ model. Using the best fitting parameters given in [1] we calculated the two $\Delta PL(f_M)$ components, as well as $|\Delta PL(f_M)|$ and the phase $\phi(f_M)$ responses shown in Fig. 2(a). It is apparent that *the TPQ model cannot describe the data in Fig. 1.* Firstly, $\Delta PL_I(f_M)$ does not change sign; this is also seen in the $\phi$ response that does not decrease beyond $\phi = \pi/2$; secondly, the two bumps in $\Delta PL_Q(f_M)$ response using Eq. (1) are not reproduced in the experimental data; and thirdly, there cannot be any dependence on the power $P$, as seen experimentally; since Eq. (1) is independent on $P$. We also tried to change the parameters $z_M$, $p_{M1}$ and $p_{M2}$ in Eq. (1) so that a zero crossing occurs in $\Delta PL_I(f_M)$ response. For the unrealistic parameters $z_M > p_{M1} + p_{M2}$ there is indeed a change in sign; however the sign change in $\Delta PL_I(f_M)$ is followed by a sign change in $\Delta PL_Q(f_M)$, in disagreement with the experimental data in Fig. 1. We conclude that the TPQ model that apparently describes $|\Delta PL(f_M)|$ response in an ad-hoc manner, is in fact inadequate to describe the complete $\Delta PL(f_M)$ response. This is *significant* since the full PLDMR response gives a clue as to the underlying physical process responsible for the resonance.

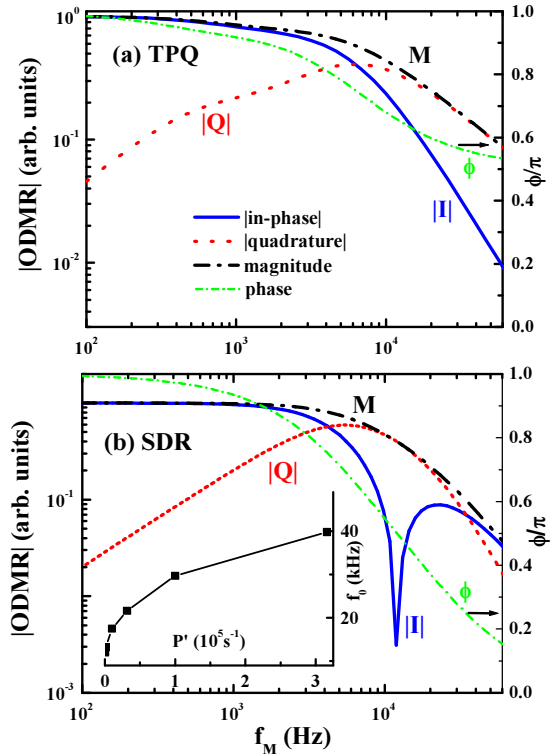

**Fig. 2:** The spin-½ PLDMR dependence on $f_M$ calculated using (a) the TPQ model (Eq. (1)), and (b) the SDR model (Eqs. (2) and (3)). The two PLDMR components are shown together with the magnitude $|\Delta PL|$ and the phase; the color codes and symbols are as in Fig. 1. The zero crossing of the in-phase component using the SDR model reproduces the data in Fig. 1. The inset in (b) shows that the calculated zero crossing frequency, $f_0$, in the SDR model increases with the microwave power similar to the data in Fig. 1 (inset). [The abscissa in the inset, $P'$, is proportional to the microwave power applied in the experiment, $P$, via: $P'(1/\text{sec}) \cong 6.7 \times 10^3\ P\ (\text{mW})$]

On the contrary, a model in which the polaron recombination is spin dependent (the SDR model [2]) describes the full PLDMR dynamic. This model, dubbed 'distant pair recombination model' has been used previously in various inorganic [3-7] and organic semiconductors [2, 8, 9]. In the SDR model polaron pairs with antiparallel spins (having population $n_1$) recombine faster than polaron pairs with parallel spins (having population $n_2$). If the polaron pairs are generated with equal initial populations, then 'spin polarization' is established by the different recombination rates of parallel and antiparallel pairs, since at steady state conditions $n_1 > n_2$. Microwave absorption reverses the spin sense of some of the polaron pairs so that at saturation $n_1 = n_2$. Therefore, the resonance conditions enhance the overall polaron recombination rate, and consequently the polaron density decreases as seen in the experiment [2]. Whether the PL increases due to reduction in polaron quenching of singlet excitons [1] or/and due to polaron pair radiative recombination [10], is a secondary question that would depend on the polymer film nanomorphology [11]; and thus has little to do with the PLDMR kinetics. The PLDMR dynamics in the SDR model is described by a pair of rate equations given by [3]:

$$dn_1/dt = G - n_1/\tau_1 - (n_1 - n_2)/2T_{sl} - (n_1 - n_2)P, \qquad (2)$$
$$dn_2/dt = G - n_2/\tau_2 - (n_2 - n_1)/2T_{sl} - (n_2 - n_1)P, \qquad (3)$$

where $G$ is the generation rate; $\tau_1$ and $\tau_2$ are the lifetimes of polaron pairs with spin antiparallel and parallel, respectively; and $T_{sl}$ is the polaron spin-lattice relaxation time. The coupled equations (2) and (3) were solved numerically, and the change $\Delta n$ in the polaron density due to the microwave power $P$ was calculated in the frequency domain. The two $\Delta n$ components, namely $\Delta n_I$ and $\Delta n_Q$, as well as $|\Delta n|$ and the phase $\phi$ were obtained as a function of $f_M$; this procedure was repeated at various $P$. In addition, an analytical approximate solution to equations close in form to Eqs. (2) and (3) also gives results similar to our numerical solution [12]. A typical ODMR $f_M$-response based on the numerical calculations of Eqs. (2) and (3) is shown in Fig. 2(b); the $f_M$ dynamics was obtained with the parameters: $\tau_1 = 14$ μsec; $\tau_2 = 60$ μsec; and $T_{sl} = 10$ μsec. In contrast to the TPQ model, it is seen that the elegant SDR model excellently describes all the PLDMR experimental response features. Firstly, the in-phase ODMR component correctly changes sign at $f_0$, following by $\phi$ passing the value $\phi = \pi/2$; secondly, the quadrature ODMR component is rather smooth and does not change sign; and thirdly, the calculations reproduce the increase of $f_0$ with $P$ (Fig. 2(b) inset). The change in sign of the in-phase ODMR is quite natural in the SDR model and does not depend on the parameters used; it in fact shows that the two spin states ($n_1$ and $n_2$) involved in the resonance have different recombination rates [13]. We therefore conclude that the SDR model is capable of describing the PLDMR dynamics in full, whereas the TPQ model does not.

Next, we studied the PLDMR and PA dynamics as a function of temperature, $\theta$. Fig. 3(a) [left panel] shows the spin-½ and spin-1 PLDMR of polarons and TE, respectively, at various temperatures. Whereas the spin-1 PLDMR sharply decreases with $\theta$ indicating that the TE spin-lattice relaxation rate dramatically increases with $\theta$; the spin-½ PLDMR hardly changes with θ. This shows that (i) at high temperatures the TE do not conserve spins, and thus cannot participate in spin-dependent collisions between TE and polarons, as required by the TPQ model; and (ii) that the dramatic increase in TE spin-lattice relaxation rate with θ has no influence over the spin-½ PLDMR resonance; the polaron and TE spin dynamics are simply *not correlated*, in contrast to the TPQ model [1].

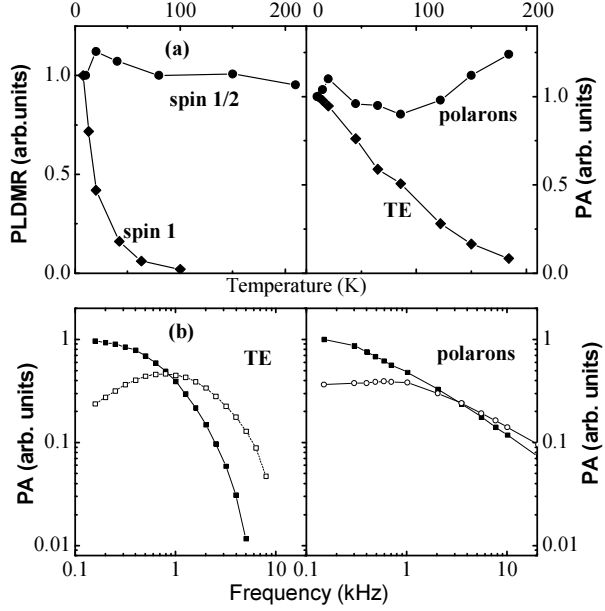

**Fig. 3:** (a) [left panel] The spin ½ PLDMR (circles), and spin-1 (squares) resonances vs. temperature; [right panel] the PA of polarons (circles) and triplet excitons (TE; squares) vs. temperature. (b) The PA dependence on the laser modulation frequency, $f_L$ for the TE [left panel] and polarons [right panel]. Both the in-phase (full symbols) and the quadrature (empty symbols) components are shown.

Fig. 3(a) [right panel] shows the PA temperature dependence of TE measured at 1.35 eV, and polarons measured at 0.4 eV [14]. In agreement with the spin ½ PLDMR, the polaron PA hardly changes with θ. By contrast, the triplet PA decreases with θ by more than an order of magnitude up to 200 K. However, the decrease in TE density with θ apparently does not have any influence on the spin-½ PLDMR resonance (Fig. 3(a), left panel), or polaron PA response; in contrast to the conditions stated above for the TPQ model [1].

Finally, we also examined the PA dynamics at low temperatures. Fig. 3(b) shows the two components of the PA vs. laser modulation frequency ($f_L$) for TE [left panel] and polarons [right panel] measured at 1.35 eV and 0.4 eV, respectively. As seen in the left panel the TE recombination kinetics may be described by a single time constant. From the crossover of the two PA components at $f_L \approx 800$ Hz, where $\omega\tau \approx 1$ [15] ($\omega = 2\pi f_L$ and $\tau$ is the TE lifetime) we get $\tau \approx 200$ μsec. This is about one order of magnitude longer than τ = 25 μsec extracted for the TE using the TPQ model for the PLDMR and PA dynamics in [1]. This shows that the TPQ model disagrees with the data. Moreover, as seen in Fig. 3(b) [right panel] the polaron recombination kinetics is *dispersive* [15], and thus cannot be described by a two-lifetime response, namely paired and unpaired polarons, as attempted in [1].

In conclusion, by measuring the full dynamics of the spin-½ and spin-1 PLDMR as a function of microwave power and temperature, together with the PA dynamics vs. temperature, we show that the TPQ model is irrelevant for describing the PLDMR and PA responses in MEH-PPV films. In contrast, we show that a competing model, namely the spin dependent recombination of polarons, which has been extensively used in previous publications, describes well the whole body of experimental results, and in particular the spin-½ PLDMR bizarre frequency response measured here.

We acknowledge fruitful discussions with Prof. M. Wohlgenannt, and thank the NSF support through DMR grant 05-03172 at the University of Utah. EE acknowledges the support of the Israel Science Foundation (735/04).